\documentclass[%
reprint,
 amsmath,amssymb,
 aps,
]{revtex4-2}

\usepackage{graphicx}
\usepackage{dcolumn}
\usepackage{bm}

\usepackage{xcolor}
\usepackage{comment}
\usepackage{mwe}
\usepackage{xspace}
\newcommand{\ndfeb}{Nd$_2$Fe$_{14}$B}
\newcommand{\micron}{$\mu$m}
\newcommand{\etal}{\mbox{et~al.}}

\begin{document}

\title{3D Imaging of Magnetic Domains in \ndfeb\ using Scanning Hard X-Ray Nanotomography}

\author{Srutarshi Banerjee$^a$}\email{sruban@anl.gov}
\author{Doga Gursoy$^a$}
\author{Junjing Deng$^a$}
\author{Maik Kahnt$^b$}
\author{Matthew Kramer$^c$}
\author{Matthew Lynn$^c$}
\author{Daniel Haskel$^a$}
\author{J\"org Strempfer$^a$}\email{strempfer@anl.gov}

\affiliation{%
$^a$X-Ray Science Division, Argonne National Laboratory, Lemont, IL - 60439, USA
$^b$ MAX IV Laboratory, Lund University, 22100 Lund, Sweden
$^c$ AMES National Laboratory, Ames, IA - 50011, USA}%

\date{\today}
\begin{abstract}
Nanoscale structural and electronic heterogeneities are prevalent in condensed matter physics. Investigating these heterogeneities in three dimensions (3D) has become an important task for understanding their material properties. To provide a tool to unravel the connection between nanoscale heterogeneity and macroscopic emergent properties in magnetic materials, scanning transmission X-ray microscopy (STXM) is combined with X-ray magnetic circular dichroism (XMCD). A vector tomography algorithm has been developed to reconstruct the full 3D magnetic vector field without any prior assumptions or knowledge. 2D STXM projections of single crystalline \ndfeb\ pillars are recorded for two different sample tilt angles while rotating around the vertical tomographic axis using $120$ nm X-ray beams with left and right circular polarization. Image alignment and iterative registration has been implemented, based on the 2D STXM projections for the two tilts. Dichroic projections obtained from difference images are used for the tomographic reconstruction to obtain the 3D magnetization distribution at the nanoscale.  
\end{abstract}

\maketitle

\section{\label{sec:intro}Introduction}

Magnetic materials play an important role in modern day industrial applications, from electric motors to sensors to high density data storage and energy harvesting. The details of a material internal magnetization structure dictates several interesting properties. For example, electric motor cores are influenced by the presence of single domain walls and their displacements \cite{arrott2019micromagnetism}. Internal magnetic structure is dictated by intrinsic properties such as lattice symmetry, spin-orbit coupling, exchange and dipolar interactions, as well as extrinsic properties such as structural defects acting as pinning centers, resulting in complex textures that affect macroscopic properties such as coercivity and remanence \cite{gutfleisch2011magnetic}.

In order to understand the behavior of magnetic systems, the determination of the internal magnetic domain structure is essential. In bulk magnets, probing these magnetic structures has been historically difficult, as they rely on indirect probing techniques. Tomographic techniques have been explored recently. In the first magnetic tomography experiments, neutron imaging was used to visualize internal magnetic fields \cite{kardjilov2008three} and magnetic domain walls \cite{manke2010three} in bulk magnetic systems with a spatial resolution of a few tens to hundreds of micrometers. Until recently, the investigation of smaller internal magnetic structures was limited to thin films and nanostructures. Photoemission electron microscopy (PEEM) as a surface sensitive probe as well as XMCD in the soft X-ray range in transmission \cite{streubel2015retrieving, blanco2015nanoscale} and Lorentz transmission electron microscopy (TEM) in combination with tomography  \cite{phatak2010three, phatak2014visualization, tanigaki2015three} are widely used. Film thicknesses were mostly limited to not more than $200$ nm in order to probe thin-film magnetism in transmission allowing characterization of relatively thin 3D magnetic structures \cite{streubel2016magnetism, fernandez2017three}. The ability to probe actual 3D nanoscale magnetic structures in the bulk of magnets opens up a new dimension for exploration of a wide range of emerging properties and phenomena such as magneto-chiral effects, \cite{hertel2013curvature, yan2012chiral}, complex magnetic configurations \cite{donnelly2017three}, high domain wall velocities \cite{hertel2016ultrafast, yan2011fast}, and asymmetric spin wave dispersion \cite{otalora2016curvature}. Dynamics of nanoscale magnetic textures has also been explored with soft x-ray PEEM and STXM using XMCD contrast \cite{schaffers2019extracting, pile2020direct}. In the last few years, hard X-ray dichroic ptychographic-tomography has been developed which offers high spatial resolution and large penetration depth and can also be adapted to thin films, allowing the investigation of a number of extended magnetic systems \cite{donnelly2017three, donnelly2018tomographic, donnelly2020time, donnelly2022complex} and thus accessing the varying magnetic structures in the sample bulk of the inherently three dimensional structures, including their dynamics \cite{donnelly_dynamics}. At the same time, hard X-ray dichroic tomography was used to obtain 2D magnetization maps by Suzuki \etal\ \cite{suzuki2018three} which then lead to the visualization of skyrmions in 3D \cite{seki2021} and the investigation of magnetization dynamics in \ndfeb\ in applied magnetic fields \cite{takeuchi2022}.

The main challenges in the development of X-ray based imaging techniques for the investigation of 3D magnetization in thick samples at the nanoscale are the following. First, limitations in the positional accuracy of the sample illuminated region during sample scanning causes misalignment of the 2D projections which leads to incorrect reconstruction of the final 3D magnetization structures of the material. Second, availability of appropriate tomographic reconstruction algorithms for recovering all three components of the magnetic vector field. Traditional tomography retrieves a scalar value at each voxel. However, for magnetic vector tomography, three orthogonal components of magnetization need to be reconstructed for each voxel. In literature, this has been achieved through using additional constraints. For instance in electron tomography, by using Maxwell's equations, the complexity can be reduced \cite{phatak2010three, phatak2008vector}. For experiments with soft X-rays, prior knowledge of the magnetic material has been incorporated \cite{streubel2015retrieving, streubel2016magnetism}. The angular dependence of the magnetic signal has also been exploited for retrieving the 3D magnetic vector field \cite{blanco2015nanoscale}. While these experiments were done in the soft X-ray range, determination of 3D magnetization in the hard X-ray regime was pioneered by Donnelly \mbox{et~al.} using 2D ptychographic projections \cite{donnelly2017three}. Due to the high transmission of the X-rays in this case the acquisition of 2D projections for a large number of rotation angles for two different sample tilts allows reconstruction of the 3D magnetic vector field by solving a set of simultaneous equations. By combining all tomographic projections it is possible to recover the three magnetization components in a single reconstruction algorithm \cite{donnelly2018tomographic}. 

In this paper, we develop an open-source end-to-end X-ray imaging algorithm to reconstruct the three components of the magnetization vector field in 3D for magnetic cylinders of a few micron diameter. Our entire imaging pipeline has been developed in Python and is available on Github \cite{gitlink}. The 2D projection images are obtained from STXM with circular dichroic contrast. These projections are aligned and registered based on a joint iterative reconstruction and reprojection method \cite{gursoy2017rapid}, which uses the principle of tomographic consistency. The magnetization is reconstructed in two perpendicular directions in 2D for the first tilt orientation of the sample. This is followed by a 3D rotation of the reconstructed object along with the magnetization to align it with the second tilt, and a subsequent reconstruction of the 3D magnetization. We demonstrate the applicability of our algorithm for two cylindrical pillars of single crystalline \ndfeb\ samples, a material which makes the basis for the best permanent magnet currently available. We retrieve the 3D magnetization vector fields from the pillars at a resolution close to $120$ nm, which corresponds to the X-ray beam probe size used in the STXM experiments.

\section{\label{sec:expt}Experimental Setup}
\begin{figure}
\centering
    \includegraphics[width=0.95\linewidth]{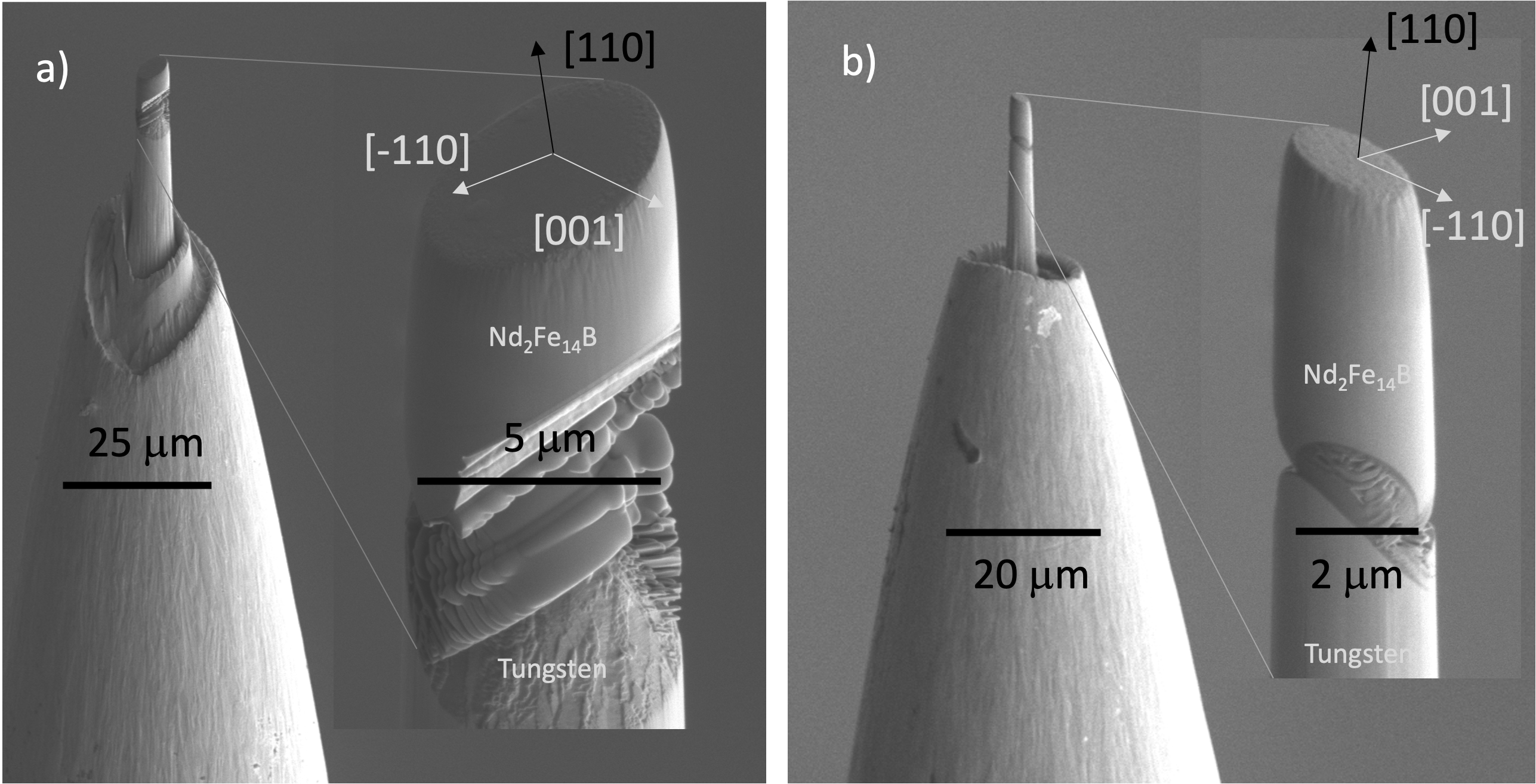}
    \caption{Electron microscope images of the (a) 5.4 \micron\ and (b) 2.1 \micron\ \ndfeb\ single crystal pillars mounted on a tungsten tip. The zoomed in images show the approximate crystal orientation with the flat top facet normal aligned close to the [110] crystallographic axis. The [-110] and [001] directions were inferred from experimental results assuming moments are oriented predominantly along the [001] easy axis.}
\label{sample_zo_zi}
\end{figure}

Two single crystal \ndfeb\ cylinders of diameters $5.4$ $\mu$m and $2.1$ $\mu$m were fabricated from flux grown crystals at Ames Laboratory \cite{wang1998} using focused ion beam (FIB) milling. A large single  crystal of approximate size of $1 \times 0.5 \times 0.5$ mm$^3$, was aligned so the sample’s long axis is the [110] direction, and the [001] is orthogonal to the long axis. A rectilinear sample about 20 µm long and about 5 µm wide was extracted from the bulk crystal using Ga ion milling. The sample was then welded to a tungsten needle and further milled to form the cylindrical sample, Fig. \ref{sample_zo_zi}(a). The procedure was repeated for the 2.1 $\mu$m sample, Fig. \ref{sample_zo_zi}(b). The normal of the inclined, flat facets at the tip are closely aligned with the [110] crystallographic axis of the sample. The [001] direction, which lies close to the radial plane, is initially unknown but can be inferred from the magnetization direction within the individual magnetic domains since \ndfeb\ is ferromagnetic and has an exceptionally high uniaxial magneto crystalline anisotropy along [001] at ambient temperature \cite{sagawa1985}.

The STXM experiments were performed at the diffraction endstation \cite{carbone2022} of the NanoMAX beamline \cite{nanomax} at the MAX IV Synchrotron, Lund, Sweden. A diamond phase plate with a thickness of $500$ $\mu$m was used to generate circularly polarized X-rays. The monochromatic circularly polarized X-ray beam was focused by Kirkpatrick-Baez (KB) mirrors resulting in a focus size of $120$ nm at the sample position \cite{bjoerling2020, kahnt2022complete}. The sample was mounted on a holder on top of a vertical rotation axis and a 2D scanner. Two different mounting holes in the sample holder allowed mounting the sample with the cylindrical axis along the vertical direction and at $30 ^{\circ}$ from it. The transmitted X-rays were detected with an Eiger2 X 1M detector \cite{donath2023}. The X-ray energy was tuned to $2$ eV above the Nd L$_2$ edge at $6727$ eV in order to obtain maximum dichroic absorption contrast. 2D maps of the transmitted beam intensity were obtained using fly scanning vertically with an exposure time of $25$ ms per point and an effective step size of $50$ nm $\times$ $50$ nm for the $2.1$ $\mu$m pillar and $100$ nm $\times$ $100$ nm for the $5.4$ $\mu$m pillar. Projections were acquired in an angular range of $360 ^{\circ}$ with step size of $1 ^{\circ}$ for incident left ($C_L$) and right ($C_R$) circularly polarized X-rays for both tilt angles while polarization was switched between $C_L$ and $C_R$ at each angular position. The 2D projections were used for the vector tomographic reconstruction of the 3D magnetic domains of the sample. 

\begin{figure}
  \centering
  \includegraphics[width=0.95\linewidth]{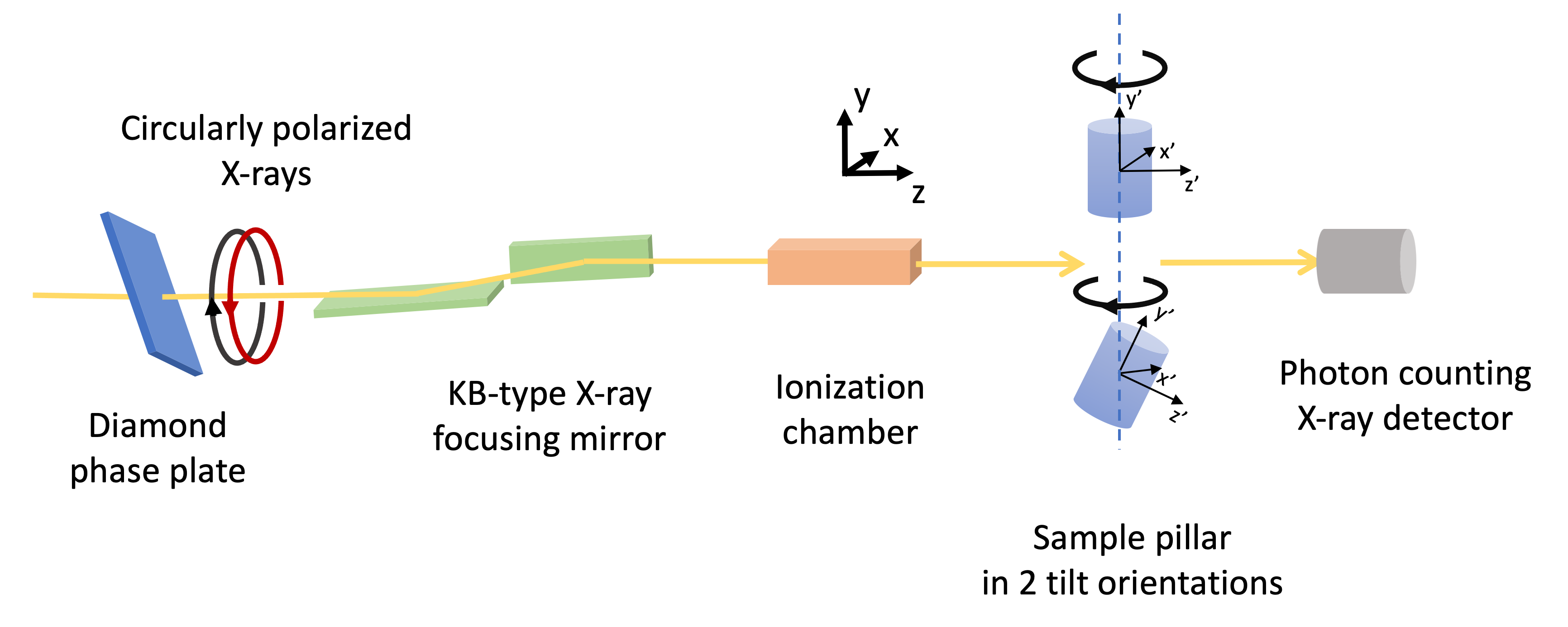}
  \caption{Schematic of the experimental setup for 3D magnetic tomography based on dichroic STXM with definition of laboratory and sample coordinate systems.}
\label{expt_setup}
\end{figure}

For magnetic contrast, XMCD is exploited, which has the maximum signal for magnetic moment parallel to the X-ray propagation direction and zero when the magnetic moment is perpendicular  \cite{siegmann2006magnetism}. The XMCD signal has low magnitude with respect to the relatively noisy background of the acquired data, and hence the signal-to-noise ratio (SNR) is low. For each angular position, the integral of the magnetic component parallel to the X-ray beam is measured in a single projection. Thus, rotating the object around the vertical \emph{y}-axis perpendicular to the X-ray direction along \emph{z} only probes the magnetization components $m_{z'}$ and $m_{x'}$ in the plane perpendicular to the rotation axis \emph{y}. In order to reconstruct all three components of magnetization $(m_{x'}, m_{y'}, m_{z'})$, dual axis tomography is performed by recording the projection data in addition, with the sample tilted at an angle of $30^{\circ}$ with respect to the rotation axis \emph{y}, enabling access to the $m_{y'}$ component as well. In the following, the two tilts with $0^{\circ}$ and $30^{\circ}$ inclination are referred to as tilt-1 and tilt-2, respectively.

The XMCD absorption contrast is defined as $\Delta \mu = \mu^{+}-\mu^{-}$, where $\mu^{+}=ln(I_{0}^{+}/I^{+})$ and $\mu^{-}=ln(I_{0}^{-}/I^{-})$ are the absorption coefficients determined from the incident, $I_{0}$, and transmitted X-ray intensities \cite{suzuki2018three}. The $+$ and $-$ superscripts refer to $C_R$ and $C_L$ X-ray polarization respectively. The flipping ratio is defined by $\Delta \mu/(\mu^{+}+\mu^{-})$ and was found to be 1.1\% at maximum contrast.

\section{\label{sec:image_alignment}Alignment of Projections}
The stability of the experimental setup for acquiring the projections is critical for the reconstruction of the 3D magnetization structures in high resolution, i.e. limited by beam size only. For the $120$ nm beam used in the experiment, a high positional accuracy and stability is desired. The stability may be compromised by vibrations and drifts in the experimental setup or instabilities in the X-ray beam, causing misalignment between X-ray beam, sample and detector. Imperfections of rotation stage motion (runout error/spindle error) or drifts caused by small temperature variations can contribute to deterioration of resolution. Displacements of subsequent projections ultimately result in incorrect reconstruction of the 3D magnetization structures. There are various kinds of instabilities which can be due to fast fluctuations, slow movements or shifts. Fast fluctuations lead to blurring, while slow movements can lead to image distortions or image displacement. Fast fluctuations need to be avoided since they cannot be corrected for. On the other hand, slow movements can be corrected for using registration techniques.

In our case, the acquired projections are aligned with respect to a common tomographic rotation axis. For small or absent drifts we can neglect image distortions and the object can be considered a rigid body, allowing rigid body registration using a two step approach to compensate for shifts between acquisitions. Alignment of all projections is done for each polarization and tilt position separately followed by reconstruction of the 3D objects using the aligned projections. For each polarization, the 3D reconstructed scalar object from tilt-1 is rotated and aligned to the 3D reconstructed scalar object from tilt-2. Conversely, the reconstructed scalar 3D object from tilt-2 is aligned with the reconstructed scalar 3D object from tilt-1 after rotating it back. This is described in the following subsections.

\subsection{\label{tomo_align}Alignment of the tomographic axis} 
The imperfection of the rotation stage at small length scales, often called runout error, causes a noticeable shift of the rotation axis horizontally and vertically. However, all magnetic contrast projection images in the tomogram must have a common axis of rotation in order to allow an accurate 3D reconstruction of the object at high resolution. In literature, techniques using fiducial markers \cite{cheng2014image,han2015novel}, natural features such as corners \cite{brandt2001multiphase}, Canny edge detection \cite{duan2008automatic} or capacitance sensors to measure the runout errors of rotation stage \cite{xu2014high} as well as tomographic consistency methods \cite{gursoy2017rapid,odstrvcil2019alignment,yu2018automatic} are used. Here, we use the joint reconstruction and reprojection (JRR) method \cite{gursoy2017rapid}, where the 3D object is reconstructed without considering sample jitter to obtain a low-resolution 3D volume. Subsequently, the reconstructed 3D object is projected back to the corresponding viewing angles through a reprojection process. The translation in horizontal and vertical direction is corrected by registering the image pairs at the different projection angles. Each image pair consists of the experimentally obtained projection image and the corresponding calculated reprojected image at the same projection angle. The cross-correlation between the image pairs provides us with the amount of horizontal and vertical shift to the experimentally acquired projection images to maximize this cross-correlation metric. Such shifts are obtained for all projections. Iterating over the reconstruction and reprojection steps $N$ times improves the alignment of the projections and subsequently the resolution of the reconstructed 3D object. 

The JRR algorithm is applied to all projections obtained with $C_L$ as well as $C_R$ polarization for one tilt. Projections corresponding to $C_L$ and $C_R$ are concatenated in an array, along with the corresponding projection angles. This then allows the alignment of projections corresponding to the two polarization states in a joint manner. The intermediate tomographic 3D object reconstruction during the JRR alignment is done using the popular simultaneous iterative reconstruction technique (SIRT) \cite{gilbert1972iterative, herman1976iterative, herman1976quadratic, lakshminarayanan1979methods, van1990sirt}. SIRT results in smoother reconstructions than other methods which allows for better alignment of the projections in the JRR algorithm. A similar JRR based alignment of the projections is repeated for the second tilt of the object. 

\subsection{\label{conv_tilts}Rotation of 3D Object between tilts}

\begin{figure}[h!]
  \centering
  \includegraphics[width=0.75\linewidth]{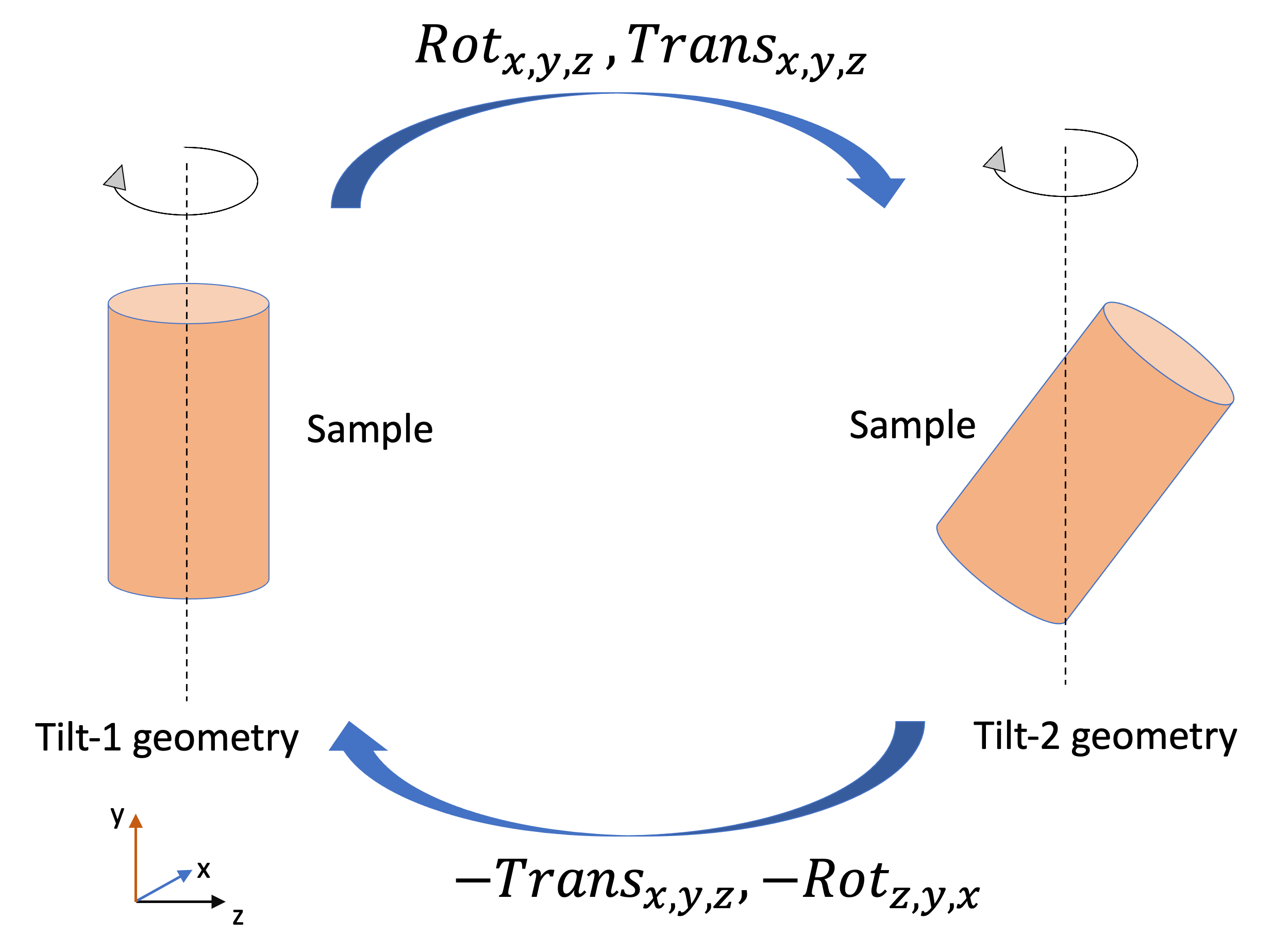}
  \caption{Rotation of the 3D reconstructed object between tilt-1 and tilt-2.}
\label{fig_rot_3D}
\end{figure}

In order to perform a full 3D magnetization vector reconstruction, the geometric transformation between the orientations of the 3D objects reconstructed at the two tilt angles must be known (see Fig. \ref{fig_rot_3D}). It is based on an initial guess of the transformation and subsequent further refinement using a conjugate gradient descent-based algorithm which searches for the perfect alignment between the original 3D object in the tilt-1 orientation and the second 3D object rotated from tilt-2 into tilt-1 orientation. The aligned projections (see section \ref{tomo_align}) corresponding to polarization states $C_L$ or $C_R$, are used to perform a scalar 3D reconstruction of the sample. The 3D object is reconstructed at tilt-1 and tilt-2 from the acquired projections using the Gridrec algorithm which is a Fourier grid reconstruction algorithm \cite{dowd1999developments, rivers2012tomorecon}. Gridrec provides us with sharper edges and boundaries of the 3D reconstructed object as it makes use of a gridding method for resampling the Fourier space from polar to Cartesian coordinates offering both computational efficiency and negligible artifacts \cite{marone2012regridding}. This enables accurate alignment of the 3D object at two different tilt orientations. Fig. \ref{fig_rot_3D} shows the rotation $Rot_{xyz}$ around $x$, $y$ and $z$ in sequence followed by translation $Trans_{xyz}$ along $x$, $y$ and $z$ in order to transform the 3D reconstructed object from the tilt-1 into the tilt-2 orientation. On the other hand in order to re-orient the same 3D object from tilt-2 to tilt-1, the object must be translated by $-Trans_{xyz}$ along $x$, $y$ and $z$ followed by a $-Rot_{zyx}$ rotation around the $z$, $y$ and $x$ axes in sequence.
Based on the experimental setup for the \ndfeb\ samples, we re-orient the 3D object from tilt-1 to tilt-2 by rotating the 3D reconstructed object around the $x$ axis, followed by rotation around the $y$ and $x$ axis successively, followed by translations of along $x$, $y$ and $z$ respectively. In order to go back to tilt-1, we apply the reverse steps for the translation followed by rotations along the same axes sequentially with the same translation and rotation parameters.

\section{\label{sec:3D_magn}Reconstruction of 3D Magnetization using Vector Tomography}
\begin{figure}
\centering
     \includegraphics[width=0.95\linewidth]{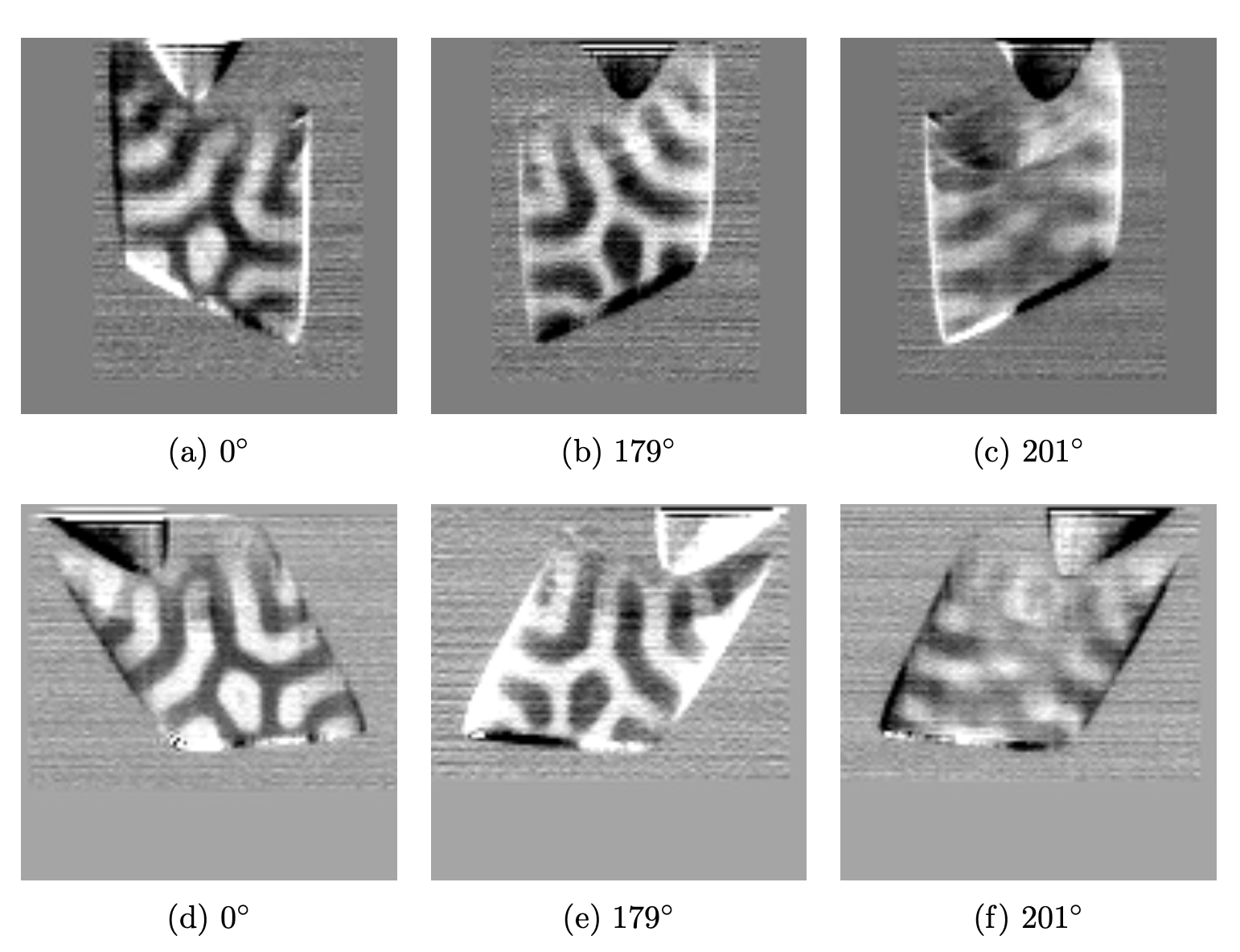}
     \caption{Cylindrical projections of the sample pillar for tilt-1 (top row) and tilt-2 (bottom row) for the same projection of magnetic domains. (b) and (e) show the sample rotated by $179^{\circ}$ with respect to (a) and (d), respectively. (c) and (f) show the projection for rotation of the object by $201^{\circ}$ with respect to (a) and (d), respectively.}
\label{Tilt_0_1 projections}
\end{figure}

After aligning the projections and the 3D reconstructed object corresponding to the two tilts as mentioned in Section \ref{sec:image_alignment} we reconstruct the 3D magnetic domains using vector tomography. We describe the magnetization using the local coordinate system $(x^{\prime}, y^{\prime}, z^{\prime})$ for the 3D cylinder. The propagation direction of the circularly polarized X-rays is along the \emph{z}-axis with the tomographic rotation axis, \emph{y}, as shown in Fig. \ref{expt_setup}. The magnetic signal measured using XMCD is proportional to the magnetization component parallel to the X-ray beam along \emph{z}. Thus, with the 3D object in tilt-1 orientation and the object being rotated around \emph{y}, initially only the magnetization components $m_{x'}$ and $m_{z'}$ are captured in the projections, as shown in Fig. \ref{Tilt_0_1 projections} (top row). On the other hand, with the 3D object in tilt-2 orientation and the object being rotated around \emph{y}-axis, the contribution of the magnetization component $m_{y'}$ as shown in Fig. \ref{Tilt_0_1 projections} (bottom row) is also captured. 

\begin{figure*}
\centering
     \includegraphics[width=0.9\linewidth]{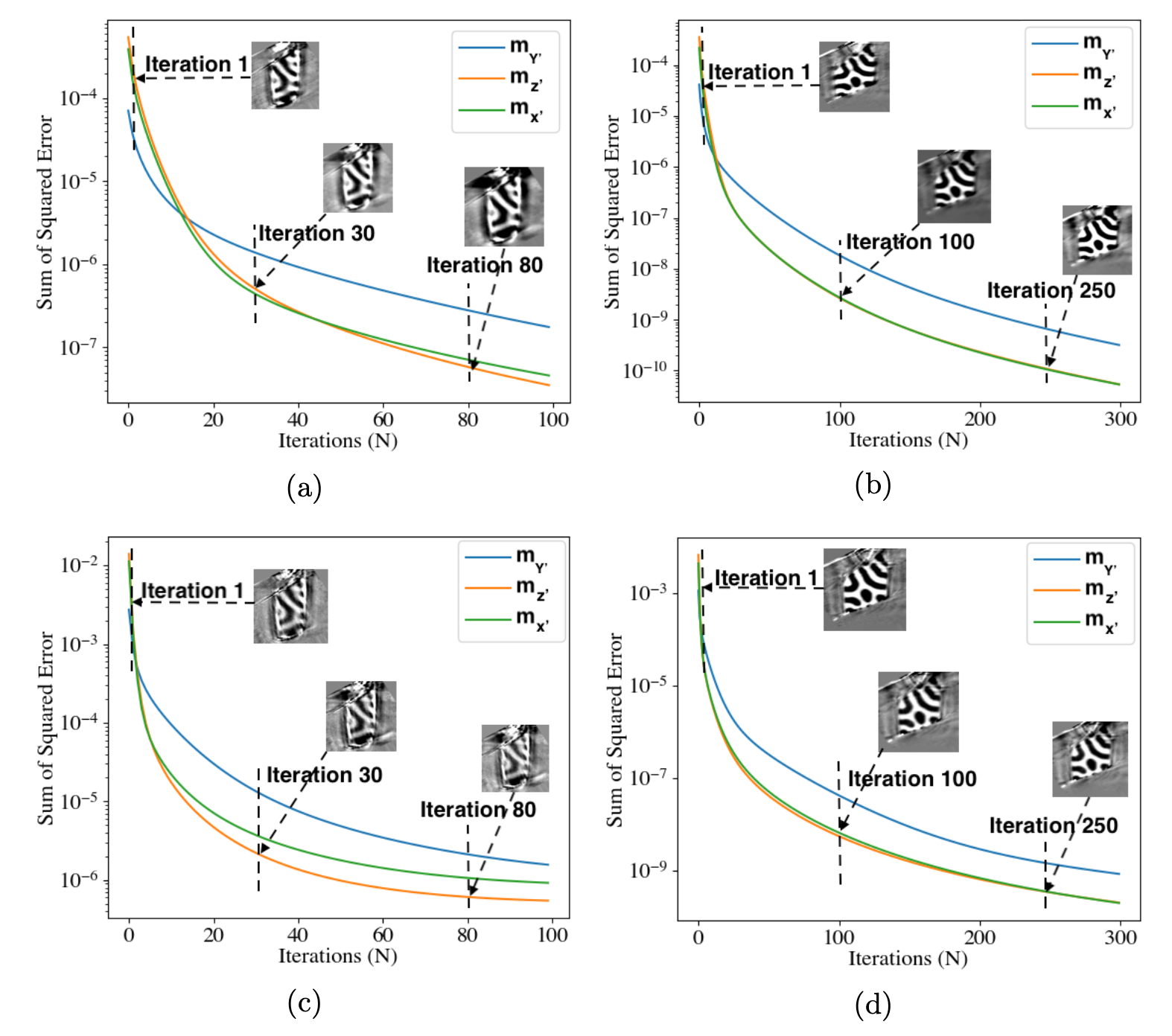}
     \caption{Sum of squared error (SSE) over iterations (N) (left column: $2.1$ $\mu$m diameter sample, right column: $5.4$ $\mu$m diameter sample), (top row: reconstruction done with one iteration for tilt-1 followed by one iteration for tilt-2, bottom row: reconstruction done with ten iterations for tilt-1 followed by ten iterations for tilt-2). The images correspond to a vertical slice at the middle of the sample for $m_{z'}$.}
\label{SSE_Iter_sample1}
\end{figure*}
The columns of Fig. \ref{Tilt_0_1 projections} show the projections for the three projection angles $0^\circ, 179^\circ, 201^\circ$ for the same projected magnetic domains corresponding to tilt-1 (top row) and tilt-2 (bottom row). From these two tilt orientations of the 3D object, the entire magnetization $(m_{x'}, m_{y'}, m_{z'})$ can be reconstructed using the SIRT algorithm. The SIRT algorithm produces less image noise and hence smoother reconstructed 3D magnetic domains than other methods. During the 3D vector magnetization reconstruction, the projections corresponding to tilt-1 are fed into the SIRT algorithm along with the initial 3D magnetization initially set to $0$. After the magnetization vectors are reconstructed for tilt-1, they are rotated and subsequently fed into the SIRT  algorithm as initial values for vector magnetization reconstruction along with the aligned projections corresponding to the tilt-2 orientation. Once the 3D magnetization vectors are reconstructed for tilt-2, they are rotated back to the tilt-1 orientation and again used as the initial value for reconstruction. This is done iteratively N times. N is determined by the fact that there is no change in the sum of squared errors (SSE) of the reconstructed 3D magnetic domains, $m_{x'}$, $m_{y'}$ and $m_{z'}$ in the successive iterations. The SSE in an iteration N is given by the formula  $ \mathrm{SSE_{N}} = \Sigma_{i,j,k=1}^{M} (f_{N}(i,j,k) - f_{N-1}(i,j,k))^2$. Fig. \ref{SSE_Iter_sample1} shows the convergence of SSE between two successive iterations over the number of iterations, N. We show the reconstructed domains for both samples. For the $2.1$ $\mu$m sample, the SSE converges faster than for the thicker $5.4$ $\mu$m sample (see Fig. \ref{SSE_Iter_sample1}). Additionally, when one iteration is done for tilt-1 followed by one iteration for tilt-2 (see Fig. \ref{SSE_Iter_sample1} (a), (b)), the convergence of the SSE is better compared to ten iterations for tilt-1 followed by ten iterations for tilt-2 (see Fig. \ref{SSE_Iter_sample1} (c), (d)). $m_{y'}$ has the largest error among the reconstructed magnetic domains as the \ndfeb\ sample is tilted by $30^{\circ}$ from the vertical direction of rotation. Since the tilt angle of $30^{\circ}$ is not large, the maximum contrast from $m_{y'}$ is reduced by a factor of two compared to $m_{x'}$ and $m_{z'}$.

\begin{figure}
\centering
     \includegraphics[width=0.95\linewidth]{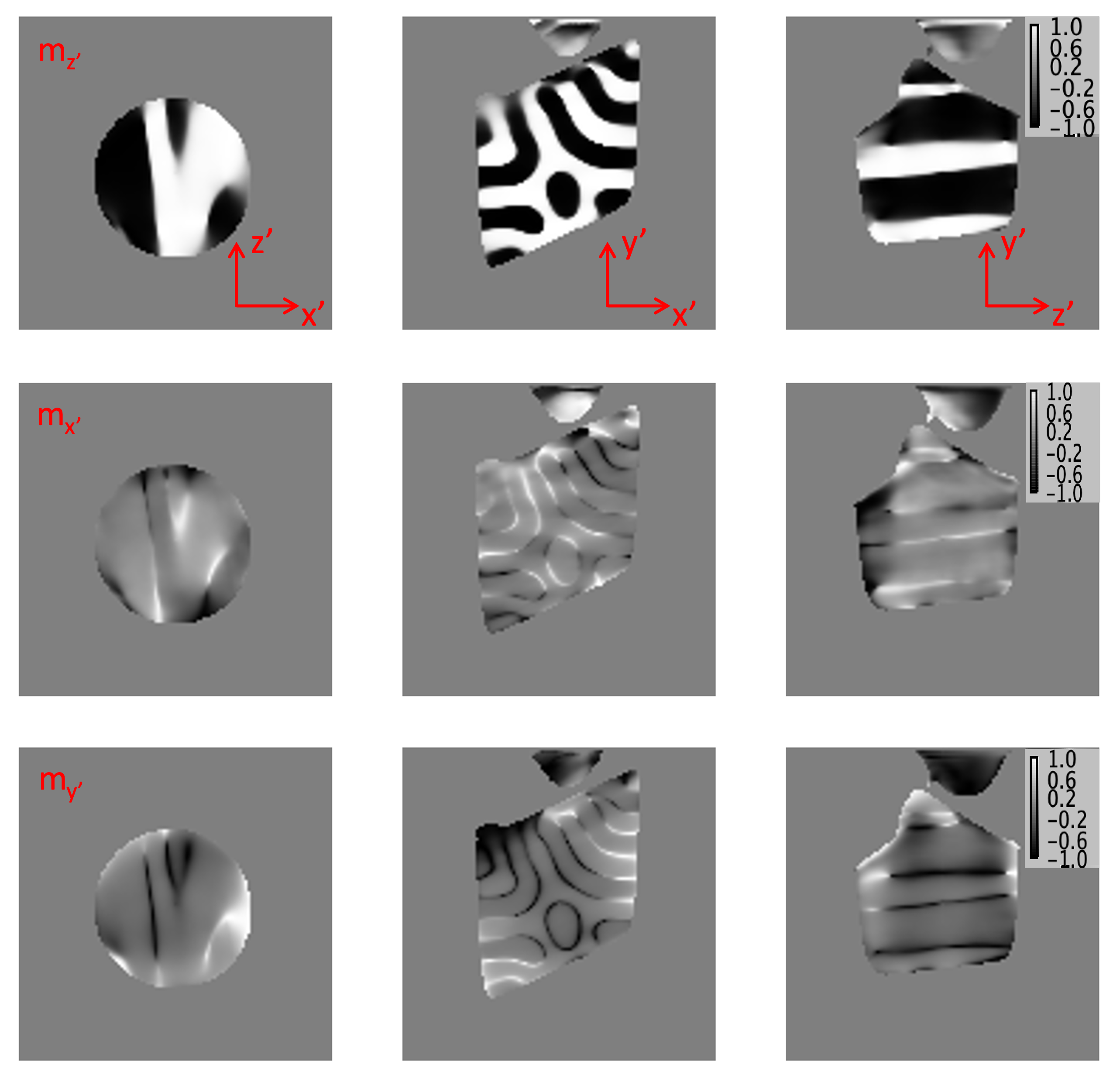}
     \caption{Horizontal (left column) and vertical (middle and right columns) slices through the center of the reconstructed $5.4$ $\mu$m diameter \ndfeb\ sample. Top row: Magnetization component $m_{z'}$ in the direction of maximum contrast. Middle row: Magnetization component $m_{x'}$. Bottom row: Magnetization component $m_{y'}$. Left column: Horizontal slice from top. Middle and right columns: Vertical slices of the sample from two perpendicular directions. The pixel minimum and maximum values are normalized to $-1$ and $1$, respectively.}
\label{my_mz_mx_reconstruction}
\end{figure}

In Fig. \ref{my_mz_mx_reconstruction}, orthogonal slices of the reconstructed 3D magnetization of the sample center are shown for components $m_{z'}$, $m_{x'}$ and $m_{y'}$. The local $z'$ axis of the \ndfeb\ sample was defined as the in-plane direction with largest magnetic contrast assumed to be aligned close to the crystallographic c-axis (easy axis). Slices through the center of the 3D magnetization reconstruction for the $m_{z'}$, $m_{x'}$ and $m_{y'}$ components are shown in the top, middle and bottom rows of Fig. \ref{my_mz_mx_reconstruction} respectively for the $5.4$ $\mu$m diameter sample. Contrast between adjacent domains for the other components $m_{x'}$ and $m_{y'}$ is low. Domain walls, although not actually well resolved are clearly visible in the slices for $m_{x'}$ and $m_{y'}$. In order to confirm if the stripes within the object slices of $m_{x'}$ and $m_{y'}$ are indeed domain walls and not artefacts due to misalignment during object registration, a vector tomographic reconstruction was also done for the sum of the $C_L$ and $C_R$ projections for both orientations with the same algorithm. No stripes were observed in this case. This affirms the stripes observed in Fig. \ref{my_mz_mx_reconstruction} are indeed domain walls.

\begin{figure}
\centering
    \includegraphics[width=\linewidth]{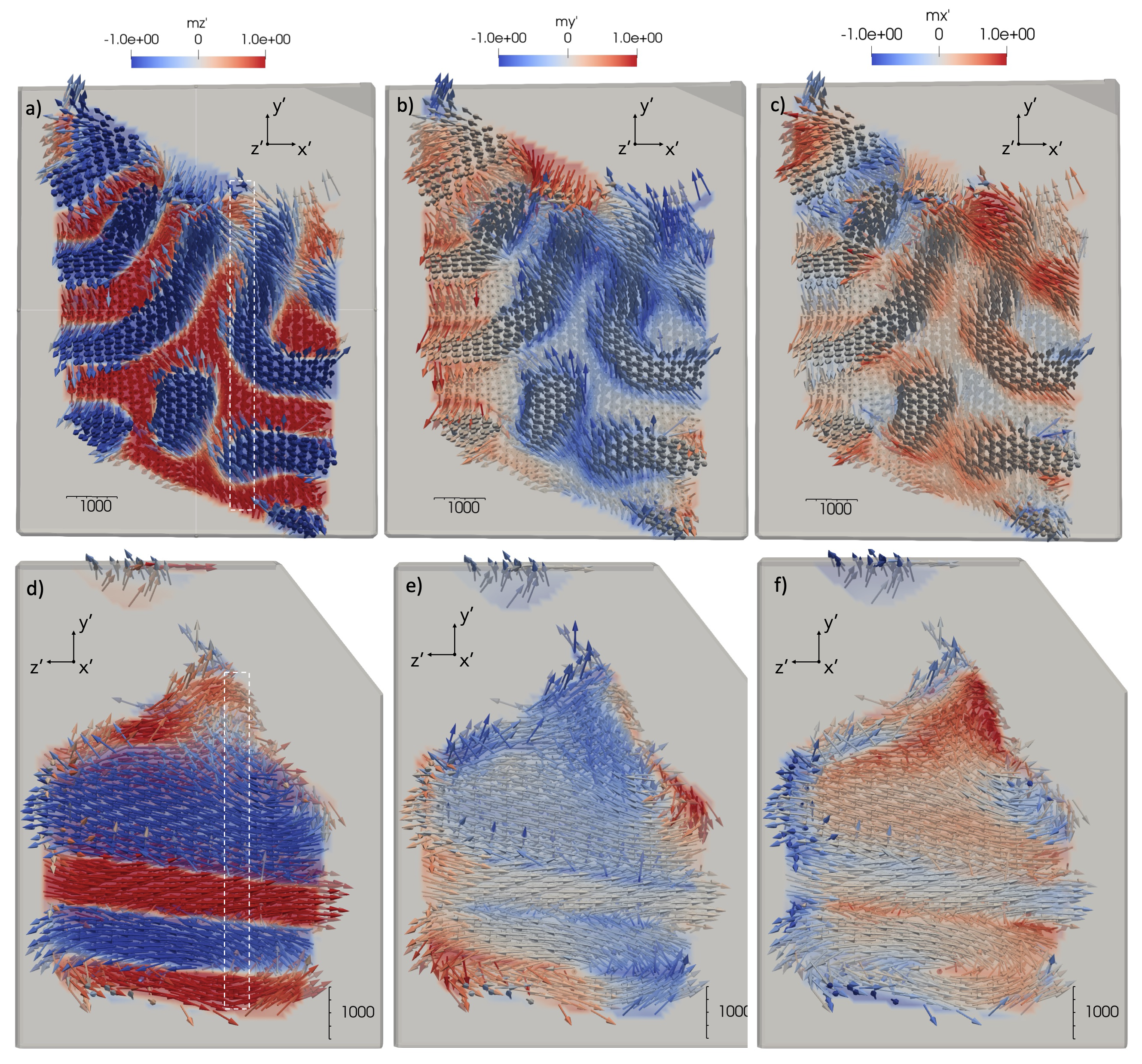}
    \caption{Slices of the reconstructed $5.4$ $\mu$m diameter sample in the orthogonal x'y' (a-c) and y'z' (d-f) planes with the moment directions visualized by vectors denoting the direction of the 3D magnetic moment. Colors denote moment components along $m_{z'}$  (a,d), $m_{y'}$ (b,e) and $m_{x'}$ (c,f), respectively, where $m_{z'}$ corresponds to the easy axis. The white rectangle in the x'y' slice denotes the location of the y'z' slice.} 
\label{5um_vol_slices}
\end{figure}

\begin{figure}
\centering
    \includegraphics[height=550pt]{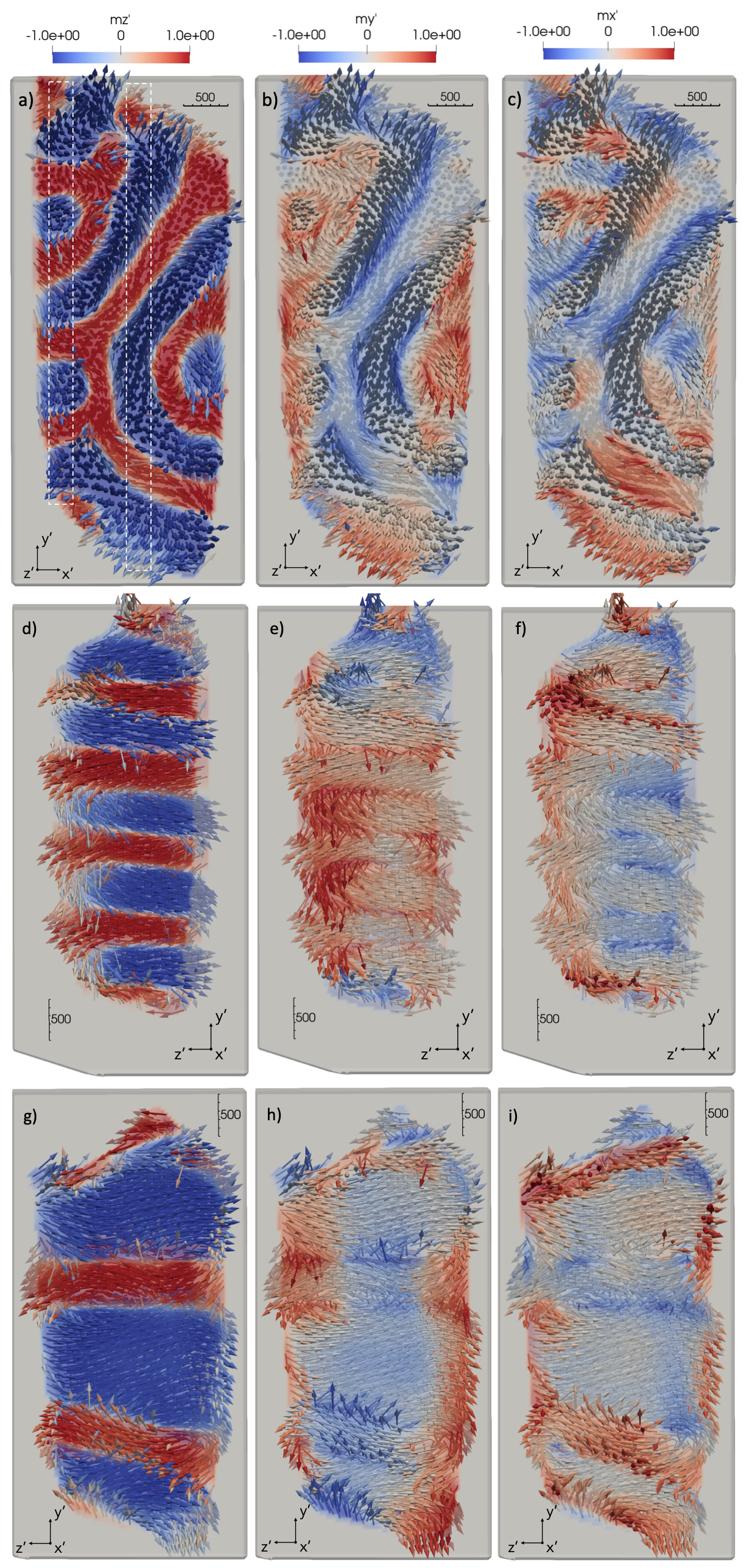}
    \caption{Slices of the reconstructed $2.1$ $\mu$m diameter sample in one orthogonal x'y' (a-c) and two y'z' planes, one close to the sample edge (d-f) and one in the sample center (g-i) with the moment directions visualized by vectors denoting the direction of the 3D magnetic moment. Colors denote moment components along $m_{z'}$  (a,d,f), $m_{y'}$ (b,e,g) and $m_{x'}$ (c,f,i), respectively, where $m_{z'}$ corresponds to the easy axis. White rectangles in the x'y' slice in (a) denote locations of y'z' slices.}
\label{2um_vol_slices}
\end{figure}

Visualization of the reconstructed 3D magnetization is done in Paraview in vector representation \cite{ahrens200536,ayachit2015paraview}. Magnetic domain vectors have been converted into 3D unit vectors. The background around the 3D object has been removed in order to visualize only the magnetic domains in the region of the 3D object of interest. The inner volume of the actual 3D object is visualized by thin, 460 nm thick orthogonal slices (5 layers). The reconstructed vector magnetization is shown in Fig. \ref{5um_vol_slices} for the $5.4$ $\mu$m diameter \ndfeb\ sample. The magnetization within the domains points in anti-parallel directions mainly along the $z'$ axis. Colour coding has been implemented for $m_{z'}$, $m_{y'}$ and $m_{x'}$ components in Fig. \ref{5um_vol_slices}(a), (b) and (c), respectively, for the x'y' slice and \ref{5um_vol_slices}(d), (e) and (f), respectively, for the y'z' slice. A similar behavior is observed for the 2 $\mu$m diameter sample in Fig. \ref{2um_vol_slices}(a)-(i). From Fig. \ref{5um_vol_slices}(d) and Fig. \ref{2um_vol_slices}(d), it can be seen that domains are extending along the direction of the magnetic moments, which is assumed to be close to the easy axis [001], but vary arbitrarily within the $x'y'$-plane. This can be seen consistently through the bulk of the pillar. As can be seen from \ref{5um_vol_slices}(a) and \ref{5um_vol_slices}(d), domain walls are mostly orthogonal to the sample surface along the x' and y' directions, whereas oriented arbitrarily within the samples. The $[110]$ direction points upwards close to the $y'$-direction. The $m_{x'}$ and especially the $m_{y'}$ components visualize the rotation of the magnetic moments in the domain walls between the ferromagnetic domains as can be seen in \ref{5um_vol_slices}(b) and (c). The domain size in the $x'y'$-plane of the $5.4$ $\mu$m sample ranges from $600$ to $800$ nm in Fig. \ref{5um_vol_slices}(a). Domains become smaller and the domain walls more extended in the $2.1$ $\mu$m sample ranging from 300 to 400 nm in Fig. \ref{2um_vol_slices}(a). The spatial resolution of the reconstructed 3D magnetic moment is determined by the size of the focused beam of $120$ nm in \emph{x} and \emph{y}.

\section{\label{sec:disc}Future Directions}
Challenges during experiments arise due to imperfections in the sample tomographic rotation stage motion and random motion in the alignment between the X-ray beam, sample and detector. This is more prominent during imaging of magnetization domain features at a spatial resolution in the order of hundred nanometers. Moreover, STXM data is acquired for $C_L$ and $C_R$ polarized X-rays for two tilt orientations with the rotation stage stability becoming crucial during the entire duration of experiment. We considered rigid body motion between the STXM projections, and aligned the projections accordingly. However, using non-rigid body motion for alignment correction would be of interest in future work to account for drifts within individual projections. The use of interferometry for registration of sample and beam position to aid in alignment is also being explored.
  
The reconstruction algorithm uses a CPU based implementation of Tomopy \cite{gursoy2014tomopy, hierro20183d}. In order to speed up the reconstructions, a GPU based implementation is desirable. A workflow can be built for experimentally acquiring the STXM projections, registering these projections based on tomographic consistency or other methods, and finally performing the 3D vector magnetization domain reconstruction in Tomopy. The algorithm developed here is released as an open source software in Github \cite{gitlink}.

With construction of the POLAR beamline as part of the Advanced Photon Source Upgrade (APS-U) underway, a dedicated capability will be available for imaging of magnetic domains in 2D and 3D \cite{strempfer2022}. The beamline will provide nanofocused beam in the 100 to 200 nm range as well as high coherent X-ray flux for ptychography experiments which will increase the spatial resolution. 

Some of these computational steps will be augmented or replaced by machine learning and other advanced algorithms. For example, the registration of the projection images and alignment of the 3D object corresponding to the two tilt geometries could be done using machine learning approaches which is one of our future research directions. Additionally, using physics-based prior knowledge from micro-magnetic simulations as an additional input to the 3D magnetization vector reconstruction algorithm is a potential new direction for speeding up the reconstruction.

\section{\label{sec:concl}Conclusion}
We developed an STXM based tomographic imaging technique that allows probing the three components of vectorial magnetization within spatially resolved magnetic domains in a magnetic material, with a resolution limited by the beam size and the accuracy of the position corrections for errors introduced by the instrument. Single crystals of \ndfeb, a material that makes the basis for strongest permanent magnets, were nano-fabricated into cylindrical samples of diameter $2.1$ $\mu$m and $5.4$ $\mu$m and used as a case study to develop the reconstruction workflow. Dichroic STXM projections are acquired by rotating the 3D object in a tomographic setup using $C_L$ and $C_R$ polarized X-rays for two sample tilt orientations. The projections are registered using a method based on tomographic consistency and the 3D reconstructed objects for the tilt orientations are aligned. A 3D magnetization vector reconstruction algorithm has been developed with the aligned STXM based projections from these two sample tilts. The different domains are clearly visible in 3D as vectors. The best estimated spatial resolution of these magnetic domains along the $\emph{x}$, $\emph{y}$ and $\emph{z}$-axes is based on the X-ray beam size of $120$ nm on the \ndfeb\ sample.

\section{\label{sec:acknw}Acknowledgements}

 We acknowledge MAX IV Laboratory for time on the NanoMAX Beamline under Proposal 20221162. Research conducted at MAX IV, a Swedish national user facility, is supported by the Swedish Research council under contract 2018-07152, the Swedish Governmental Agency for Innovation Systems under contract 2018-04969, and Formas under contract 2019-02496. This research used resources of the Advanced Photon Source, a U.S. DOE Office of Science User Facility operated for the DOE Office of Science by Argonne National Laboratory under Contract No. DE-AC02-06CH11357. Sample growth and focused ion milling was performed at Ames Laboratory which is
supported by the U.S. Department of Energy (DOE), Office of Science, Basic Energy
Sciences, Materials Science and Engineering Division. Ames Laboratory is operated for
the U.S. DOE by Iowa State University under contract No. DE-AC02-07CH11358. The
authors also acknowledge Claire Donnelly for helpful discussions, Paul Canfield for the samples and Suzanne G. E. te
Velthius for support with MOKE Microscopy.

\bibliography{apssamp}
\end{document}